\documentclass{entcs}

\usepackage{entcsmacro}
\usepackage[utf8]{inputenc}
\usepackage{mathtools}
\usepackage{microtype}
\usepackage{paralist}
\usepackage{smashtimes}
\usepackage{stmaryrd}
\usepackage[all,cmtip]{xy}

\xymatrixcolsep={1em}
\xymatrixrowsep={0.75em}

\DeclareMathOperator{\Ima}{\mathsf{Im}}
\DeclareMathOperator{\Ide}{\mathsf{Ide}}
\DeclareMathOperator{\Mo}{\mathsf{Mo}}
\DeclareMathOperator{\dom}{\mathsf{dom}}

\DeclareMathOperator{\fgraph}{\mathsf{graph}}
\DeclareMathOperator{\Ord}{\mathsf{Ord}}
\DeclareMathOperator{\IsAcq}{\mathsf{IsAcq}}
\DeclareMathOperator{\IsRel}{\mathsf{IsRel}}

\DeclareMathOperator{\RLoc}{\mathsf{RaceLocs}}

\newcommand{\rn}[1]{\textsc{(#1)}} 

\newcommand{\pto}{\rightharpoonup}
\newcommand{\pfin}{\rightharpoonup_{\mathit{fin}}}

\newcommand{\Acts}{\mathcal{A}}
\newcommand{\Ar}{{\Acts_r}}
\newcommand{\Aw}{{\Acts_w}}
\newcommand{\Af}{{\Acts_f}}

\newcommand{\VExp}{\mathbf{Exp}_\mathit{int}}
\newcommand{\BExp}{\mathbf{Exp}_\mathit{bool}}
\newcommand{\Cmd}{\mathbf{Cmd}}
\newcommand{\Prog}{\mathbf{Prog}}

\newcommand{\Ides}{\mathbf{Ide}}
\newcommand{\V}{\mathbf{V}}

\newcommand{\earead}[2]{{#2}_{#1}}
\newcommand{\etrue}{\mathbf{true}}

\newcommand{\ermw}[3]{\mathbf{rmw}(#2_{#1}; #3)}

\newcommand{\cskip}{\mathbf{skip}}
\newcommand{\cfence}[1]{\mathbf{fence}_{#1}}
\newcommand{\cif}[3]{\mathbf{if} \ #1\ \mathbf{then}\ #2\ %
  \mathbf{else}\ #3}
\newcommand{\cawrite}[3]{#2 \mathrel{\coloneqq}_{#1} #3}

\newcommand{\cifthen}[2]{\mathbf{if}\ #1\ \mathbf{then}\ #2}
\newcommand{\cwhile}[2]{\mathbf{while}\ #1\ \mathbf{do}\ #2}
\newcommand{\parop}{\mathbin{{\parallel}}}
\newcommand{\seqop}{\mathbin{{;}}}
\newcommand{\relop}{\mathbin{\raisebox{-2pt}{\vdots}}}

\newcommand{\crmw}[3]{\mathbf{rmw}_{#1}(#2; #3)}
\newcommand{\clocal}[3]{\mathbf{local}\ {#1}_\nonat = {#2}\ \mathbf{in}\ {#3}}

\newcommand{\scons}[2]{#1 \Uparrow #2}
\newcommand{\mstate}[1]{[\,#1\,]}
\newcommand{\emptystate}{\left[\,\right]}
\newcommand{\supdm}[1]{\mstate{#1}}
\newcommand{\supd}[2]{\supdm{\penalty 10000 #1 \penalty 10000 \mid \penalty 10000 #2\penalty 10000}}

\newcommand{\scomb}[4]{(#1 \sqcup (#3 \setminus \dom #2), \supd{#2}{#4})}
\newcommand{\rscomb}[3]{(#1 \sqcup (#3 \setminus \dom #2), \top)}

\newcommand{\atomic}{\mathit{at}}
\newcommand{\nonat}{\mathit{na}}
\newcommand{\relaxed}{\mathit{rlx}}
\newcommand{\release}{\mathit{rel}}
\newcommand{\acquire}{\mathit{acq}}
\newcommand{\acqrel}{\mathit{ar}}
\newcommand{\seqcons}{\mathit{sc}}

\newcommand{\ara}[3]{#2\,{=}_{#1}\,#3}
\newcommand{\awa}[3]{#2\,{\coloneqq}_{#1}\,#3}

\newcommand{\armw}[4]{\mathrm{rmw}_{#1}(#2;#3;#4)}
\newcommand{\afence}[1]{\mathrm{fence}_{#1}}

\newcommand{\defin}[1]{\textbf{\textsf{#1}}}

\newcommand{\vtrue}{\mathit{true}}
\newcommand{\vfalse}{\mathit{false}}
\newcommand{\Bools}{\mathbf{Bool}}

\DeclareMathOperator{\Pom}{\mathcal{P}}

\newcommand{\Pomt}[1]{\Pom(#1)_{\vtrue}}
\newcommand{\Pomf}[1]{\Pom(#1)_{\vfalse}}
\newcommand{\Poms}{\mathbf{Pom}}

\newcommand{\co}{\mathrel{\mathbf{co}}}
\newcommand{\dr}{\mathrel{\mathbf{dr}}}
\newcommand{\rsc}{\mathrel{\mathbf{rsc}}}
\newcommand{\rc}{\mathrel{\mathbf{rc}}}
\newcommand{\prefix}[3]{#1 \lhd_{#2} #3}
\newcommand{\PRestr}[2]{#1 \restriction #2}
\newcommand{\SExec}[2]{\mathrm{SeqExec}_{#1}(#2)}
\newcommand{\emptypset}{\mathbf{0}}
\newcommand{\Pequiv}{\equiv_{\Pom}}

\newcommand{\footp}[1]{\llbracket\,#1\,\rrbracket}

\newcommand{\Exec}{\mathcal{E}}

\let\emptyset\varnothing
\def\eg{{e.g.\@}} 
\def\cf{{\em cf.\@}}
\def\ie{{i.e.\@}}
\def\ea{{et~al.\@}}

\usepackage{todonotes}

\def\hdrdate{\leavevmode\hbox{\the\year-\twodigits\month-\twodigits\day}}
\def\twodigits#1{\ifnum#1<10 0\fi\the#1}
\def\lastname{Kavanagh and Brookes}

\begin{document}
\begin{frontmatter}
  \title{A denotational account of C11-style memory}
  \author{Ryan Kavanagh\thanksref{rkavanagh@cs.cmu.edu}}
  \author{Stephen Brookes\thanksref{brookes@cs.cmu.edu}}
  \address{Computer Science Department\\Carnegie Mellon University\\Pittsburgh, PA, USA}
  \thanks[rkavanagh@cs.cmu.edu]{Email:
    \href{mailto:rkavanagh@cs.cmu.edu}{\texttt{\normalshape rkavanagh@cs.cmu.edu}}.
    Funded in part by a Natural Sciences and Engineering Research Council of Canada Postgraduate Scholarship.}
  \thanks[brookes@cs.cmu.edu]{Email:
    \href{mailto:brookes@cs.cmu.edu}{\texttt{\normalshape brookes@cs.cmu.edu}}}
  \begin{abstract}
    We introduce a denotational semantic framework for shared-memory concurrent programs in a C11-style memory model.
    This denotational approach is an alternative to techniques based on ``execution graphs'' and axiomatizations, and it allows for compositional reasoning.
    Our semantics generalizes from traces (sequences of actions) to pomsets (partial orders of actions): instead of traces and interleaving, we embrace ``true'' concurrency.
    We build on techniques from our prior work that gives a denotational semantics to SPARC TSO.
    We add support for C11's wider range of memory orderings, \eg, acquire-release and relaxed, and support for local variables and various synchronization primitives, while eliminating significant amounts of technical bookkeeping.
    Our approach features two main components.
    We first give programs a syntax-directed denotation in terms of sets of pomsets of memory actions.
    We then give a race-detecting executional interpretation of pomsets using footprints and a local view of state.
  \end{abstract}
\end{frontmatter}


\section{Introduction}
\label{sec:examples}

A memory model specifies which values can be read by memory accesses.
C11-style memory models allow the programmer to specify if a given memory location should be acted on atomically or non-atomically~\cite{Lahav:2017:RSC:3062341.3062352}.
Atomic memory locations are intended to be used for inter-thread communication and synchronization.
Every atomic memory action has a programmer-chosen memory ordering tag.
The action's memory ordering specifies the visibility of actions sequenced before or after it to other actions that synchronize with it.
Intuitively, the memory ordering stipulates how the memory action can be reordered with other actions in the same thread.
Following Lahav~\ea~\cite{Lahav:2017:RSC:3062341.3062352}, we differ from C11 and do not treat unsequenced races between atomic accesses to the same location as undefined behaviour.
In contrast, multiple concurrent accesses to a non-atomic location, at least one of which is a write, constitutes a race and is regarded as undefined behaviour, because reading from a non-atomic location could retrieve a value from an intermediate state.
Memory actions on non-atomic locations can be compiled to normal memory accesses, cheaper to perform than their atomic counterparts.

We provide a denotational framework for exploring C11-style memory models.
We did not set out to exactly capture any particular account of the C11 memory model for two reasons.
First, because the literature presents many different accounts of the C11 memory model (\eg, \cite{Batty:2016:OSA:2837614.2837637,Lahav:2017:RSC:3062341.3062352,Vafeiadis:2015:CCO:2676726.2676995}), each addressing various shortcomings, we believe it is better to develop a generic framework in which we can study various formulations.
Second, C11 has some features, such as \textit{consume} accesses, that we deem are premature or introduce excessive complexity with little gain.

Our framework has two major components.
In Section~\ref{sec:pomset-denotations}, we give a denotational (and hence compositional) account of a C11-style memory model.
Each program is given a set of pomsets (a generalization of traces) as its denotation using various composition operators designed to capture exactly the per-thread memory reorderings permitted by the memory model.
In Section~\ref{sec:exec-interpr}, we give these pomsets an executional interpretation inductively defined on the structure of the pomset, using a local view of state.
This interpretation is carefully constructed to respect synchronization constraints provided by the memory model and it is race-detecting.

\section{An Informal Account}

Each type of memory location has an associated set of actions.
Non-atomic locations can be read from and written to.
Atomic locations can additionally be acted on with atomic read-modify-write actions.
There are memory operations that involve no locations.
For example, a fence is a special atomic memory action that acts as a barrier against reordering.
All atomic operations have an associated memory ordering tag chosen by the programmer.

The strongest ordering on atomic actions is \textit{sequential consistency}, denoted by the tag $\seqcons$.
An $\seqcons$ action cannot be reordered with any other action, and every execution induces a total order on the $\seqcons$ actions.
Though $\seqcons$ actions are expensive to implement, they allow the programmer to reason via an interleaving semantics.
All atomic memory actions can use this ordering.

The \textit{release-acquire} memory ordering paradigm gives lightweight synchronization between threads.
No memory action sequenced before a \textit{release} ($\release$) write can be reordered to after the write.
Symmetrically, no action sequenced after an \textit{acquire} ($\acquire$) read can be reordered to before the read.
The intended semantics is that any action after an acquire read that ``synchronizes with'' a release write to the same atomic location sees the effects of all actions that occurred before the write.
Fences and atomic read-modify-write actions, such as locking primitives, can use the \textit{acquire-release} ($\acqrel$) ordering.
These actions behave both as an acquire read and a release write.
A key difference from $\seqcons$ is that executions need not induce a total order on $\acqrel$ actions.

The weakest memory ordering we consider is \textit{relaxed} ($\relaxed$).
It imposes no additional constraints on reordering and only guarantees atomicity.

It is helpful to visualize the relative strength of these memory orderings using the following diagram by Lahav~\ea~\cite{Lahav:2017:RSC:3062341.3062352}:
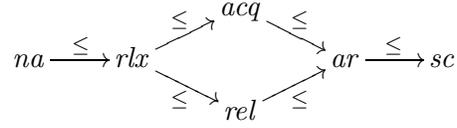
\[\xymatrixcolsep={2em}
  \xymatrix{&&\acquire \ar[dr]^{\leq}&&\\
    \nonat \ar[r]^{\leq} & \relaxed \ar[ur]^{\leq} \ar[rd]_{\leq} & & \acqrel\ar[r]^{\leq} & \seqcons\\
    && \release \ar[ur]_{\leq} & &
  }
\]

A key desideratum is that executing a single sequential thread under our memory model should produce the same result as execution without any reorderings.
This implies that at no point may we reorder memory actions to the same location within a given thread.
We also want \defin{coherence}, \ie, the property that writes to the same location appear in the same order to all threads.
Our model should further respect data dependencies: whenever we write the value of an expression to a location, any reads required to evaluate that expression must be ordered \textit{before} the write.

We illustrate these principles and the interplay between the release and acquire orderings using a simple message-passing example.
Consider executing the program
\begin{equation}
  \cawrite{\nonat}{x}{42} \seqop \cawrite{\release}{y}{1} \parop (\cwhile{\earead{\acquire}{y} = 0}{\cskip}) \seqop \cawrite{\relaxed}{z}{\earead{\nonat}{x}}\label{eq:1}
\end{equation}
from an initial state where all locations are initialized to $0$.
The fact that the write to $y$ is a release and the reads from $y$ are acquires guarantees that after the while loop terminates, the read from $x$ will see the value $42$.
We make these dependencies explicit by means of diagrams, where an arrow $a \to b$ indicates that memory action $a$ is sequenced before $b$, and ${\xymatrixcolsep={1em}\xymatrix@1{a\ar@{-->}[r] & a}}$ abbreviates $a \to a \to \cdots \to a$:
\[
  \xymatrix{
    \awa{\relaxed}{x}{42}\ar[r] & \awa{\release}{y}{1} & \ara{\acquire}{y}{0}\ar@{-->}[r] & \ara{\acquire}{y}{0}\ar[r] & \ara{\relaxed}{x}{42}\ar[r]& \awa{\relaxed}{z}{42}
  }
\]
Had all of the actions been tagged with the $\relaxed$ ordering, the compiler could have reordered the writes to $x$ and $y$ because they do not depend on each other, and then the read from $x$ could have returned 0 instead of 42, giving us
\[
  \xymatrix{
    \awa{\relaxed}{x}{42} & \awa{\relaxed}{y}{1} & \ara{\relaxed}{y}{0}\ar@{-->}[r] & \ara{\relaxed}{y}{0}\ar[r] & \ara{\relaxed}{x}{42}\ar[r]& \awa{\relaxed}{z}{42}
  }
\]
We pause to remark that the reads from $y$ are still ordered before the read from $x$: this is because there is a control-flow dependency between the read $\earead{\relaxed}{y}$ from the loop test and the write $\cawrite{\relaxed}{z}{\earead{\relaxed}{x}}$ immediately following.
Preserving control-flow dependencies will be important for eliminating various \textit{thin air} behaviours.

\section{Denotational Semantics}
\label{sec:pomset-denotations}

We make our informal account precise using a denotational semantics.
Its chief advantage is \textit{compositionality}: the meaning (or \textit{denotation}) of a program in the memory model is determined by the denotations of its subphrases.
This allows for modular reasoning and the validation of various program-level optimizations.
The denotations of programs will be order structures on syntactic objects called ``memory actions''.
These order structures, called partially-ordered multisets (pomsets), generalize traces and are an abstract description of the program's memory accesses.

We specify a simple imperative language with while loops, fences, local variables, and atomic read-modify-write actions.
These features were chosen to illustrate the principles underlying our approach, but the details of the language are not important.
We assume meta-variables for disjoint sets of identifiers $a \in \Ides_\atomic$ (atomic assignable identifiers) and $n \in \Ides_\nonat$ (non-atomic assignable identifiers) and we let $x, y$ range over the set $\Ides = \Ides_\atomic \cup \Ides_\nonat$ of all identifiers.
Finally, we let $v \in \V = \mathbb{Z}$ (integer values) and $f$ range over partial functions of type $\V \pto \V$.

The abstract syntax of our language is given by the following grammar:
\begin{align*}
  \alpha &\Coloneqq \relaxed \mid \release \mid \acquire \mid \acqrel \mid \seqcons\\
  \mu &\Coloneqq \nonat \mid \alpha\\
  e &\Coloneqq v \mid \earead{\alpha\notin\{\release,\acqrel\}}{a} \mid \earead{\nonat}{n} \mid \ermw{\alpha}{a}{f} \mid e_1 + e_2 \mid \cdots\\
  b &\Coloneqq \etrue \mid \neg b \mid b_1 \lor b_2 \mid e_1 < e_2 \mid \cdots\\
  c &\Coloneqq \cskip \mid \cawrite{\alpha\notin\{\acquire,\acqrel\}}{a}{e} \mid \cawrite{\nonat}{n}{e} \mid \cfence{>\relaxed} \mid \crmw{\alpha}{a}{f}\\
         &\;\mid \clocal{n}{v}{c} \mid c_1 \seqop c_2 \mid \cif{b}{c_1}{c_2} \mid \cwhile{b}{c}\\
  p &\Coloneqq c \mid p \parop c
\end{align*}
The meta-variables are $e \in \VExp$ (integer expressions), $b \in \BExp$ (boolean expressions), $c \in \Cmd$ (commands), and $p \in \Prog$ (programs).
We abuse notation and use subscripts such as ${>}\mu$ to mean all memory orderings $\mu'$ such that $\mu' > \mu$.
Though the phrase $\ermw{\alpha}{a}{f}$ is used both as an expression and a command, context will make its syntactic class unambiguous.
In examples, we let the associated memory order tag determine whether an identifier corresponds to an atomic or non-atomic assignable.

Our semantic clauses will transform our language's syntactic phrases into sets of memory action pomsets.
A \defin{(memory) action} $\lambda$ is a syntactic object representing an action on the store.
Memory actions are given by the following:
\begin{align*}
  \lambda &\coloneqq \delta &&(\text{no-op})\\
          &\;\mid \awa{\nonat}{n}{v} &&(\text{non-atomic write})\\
          &\;\mid \awa{\alpha}{a}{v} &&(\text{atomic write}, \alpha \in \{\relaxed, \release, \seqcons\})\\
          &\;\mid \ara{\nonat}{n}{v} &&(\text{non-atomic read})\\
          &\;\mid \ara{\alpha}{a}{v} &&(\text{atomic read}, \alpha \in \{\relaxed, \acquire, \seqcons\})\\
          &\;\mid \afence\alpha &&(\text{atomic fence}, \alpha > \relaxed)\\
          &\;\mid \armw{\alpha}{a}{v}{v'} &&(\text{atomic read-modify-write: read $v$ and write $v'$}).
\end{align*}
Let $\Acts$ be the set of all actions and let $\Mo$ and $\Ide$ be the (partial) projections from $\Acts$ to memory orderings and $\Ides$, respectively.
Given a predicate $\pi$ on $\Acts$, let $\Acts_\pi$ be the set of actions satisfying $\pi$.
For example, $\Acts_{<\mu}$ is the set of all actions $\lambda$ such that $\Mo(\lambda) < \mu$.
Special cases are the set $\Acts_\mu$ of all memory actions with memory ordering $\mu$, the set $\Ar$ of all reading actions ($\ara{\mu}{i}{v}$ and $\armw{\alpha}{a}{v}{v'}$), the set $\Aw$ of all writing actions ($\awa{\mu}{i}{v}$ and $\armw{\alpha}{a}{v}{v'}$), the set $\Af$ of fence actions, and the set $\Acts_x$ of all actions involving the identifier $x$.

A \defin{pomset} over a set of labels $L$ is a triple $(P, <, \Phi)$ where $(P, <)$ is a strict partial order satisfying the finite-height property and $\Phi : P \to L$ is a labelling function.
A partial order $(P, <)$ satisfies the \defin{finite-height property} if for all $p \in P$, the set $\{ q \in P \mid q < p \}$ is finite.
Because our pomsets describe orderings between a program's memory actions, the finite-height property implies that we have no unreachable actions in our pomset, \ie, that every action could in principle be executed.
As is typical with mathematical structures, we call a pomset by its underlying set, and given a pomset $P$, we let $<_P$ and $\Phi_P$ denote its obvious components.
We denote by $\Poms(L)$ the set of pomsets over $L$.
We typically refer to pomset elements by their labels, relying on context to disambiguate which underlying element we mean.
In fact, the underlying elements carry no meaning, and we identify pomsets $P, Q \in \Poms(L)$ whenever there exists an order-isomorphism $\Psi : (P, <_P) \to (Q, <_Q)$ respecting labels, \ie, satisfying $\Phi_Q \circ \Psi = \Phi_P$.
We can identify pomsets and labelled directed acyclic graphs, as we did in Section~\ref{sec:examples}, where we have an edge $\Phi(a) \to \Phi(b)$ if and only if $a < b$.
We say a pomset $P$ is \defin{linear} if $<_P$ is a total order.

We further identify non-empty pomsets $P, Q \in \Poms(\Acts)$ whenever there exists a non-empty pomset $R \in \Poms(\Acts)$ such that $R$ can be obtained from $P$ and from $Q$ by deleting finitely many $\delta$ actions.\footnote{Formally, the \defin{deletion} of $S \subseteq P$ from $P$ is given by $(P \setminus S, {<_P} \cap ((P \setminus S) \times (P \setminus S)), \Phi \restriction (P \setminus S))$. Deleting finitely many $\delta$ actions from $P$ means deleting a finite subset of $\Phi_P^{-1}(\delta)$ from $P$.}
This is akin to closure under stuttering and mumbling in trace semantics (\cf~\cite{BROOKES1996145}), and our semantics is well-defined relative to it.

Our semantic clauses assign sets of pomsets to syntactic phrases, and compositionality requires us to be able to compose the pomsets from subphrases to form the denotation of a phrase.
The \defin{sequential composition} $(P_1, {<_1}, \Phi_1)\seqop(P_2, {<_2}, \Phi_2)$ of pomsets $P_1$ and $P_2$ is $(P_1 \uplus P_2, {<}, \Phi_1 \uplus \Phi_2)$ when $P_1$ is finite, where $(i, p) < (j, q)$ if and only if $i = j$ and $p <_i q$, or $i < j$, and where $(\Phi_1 \uplus \Phi_2)(i,p) = \Phi_i(p)$.
Intuitively, this orders everything in $P_1$ before everything $P_2$ while preserving their internal orderings.
When $P_1$ is infinite,  $P_1 \seqop P_2 = P_1$.
The finiteness check on $P_1$ ensures $P_1\seqop P_2$ satisfies the finite-height property.
The \defin{parallel composition} $(P_1, {<_1}, \Phi_1) \parop (P_2, {<_2}, \Phi_2)$ of pomsets $P_1$ and $P_2$ is given by $(P_1 \uplus P_2, {<}, \Phi_1 \uplus \Phi_2)$ where $(i,p) < (j, q)$ if and only if $i = j$ and $p <_i q$.
It is straightforward to check that these compositions are all associative with the empty pomset $\emptypset = (\emptyset, \emptyset, \emptyset)$ as their unit.
They lift to sets of pomsets in the obvious manner.

The denotation of an integer expression is a subset of $\Poms(\Acts)\times\V$ inductively defined on the syntax of the expression:
\begin{align*}
  \Pom(v) &= \{ (\{\delta\}, v) \}\\
  \Pom(\earead{\mu}{x}) &= \{ (\{\ara{\mu}{x}{v}\},v) \mid v \in V\}\\
  \Pom(\crmw{\alpha}{a}{f}) &= \{ (\{ \armw{\alpha}{a}{v}{v'} \},v') \mid (v, v') \in \fgraph(f) \}\\
  \Pom(e_1 + e_2) &= \{ (P_1 \parop P_2, v_1 + v_2) \mid (P_i, v_i) \in \Pom(e_i) \}
\end{align*}
The $\earead{\mu}{x}$ clause has a pomset $\{\ara{\mu}{x}{v}\}$ for each possible value $v$ that could be read from $x$.
We must allow for all possible values to get compositionality: we do not know \textit{a priori} with which writes an expression may be composed, and hence do not know what values might be read from $x$.
The $\crmw{\alpha}{a}{f}$ clause is analogous and captures the atomic nature of the read-modify-write by treating it as a single memory action, rather than a sequenced read-write pair.
We indicate that we compute the $e_i$ in $e_1 + e_2$ in parallel by combining the memory actions $P_i$ with a parallel composition.

The denotation of a boolean expression is a subset of $\Poms(\Acts)\times\Bools$ and is defined analogously.
To simplify the clauses with conditionals, we introduce the helper definitions $\Pomt{b} = \{ P \mid (P, \vtrue) \in \Pom(b) \}$ and the analogous $\Pomf{b}$.

The denotation of a program $p$ is a subset of $\Poms(\Acts)$ inductively defined on its syntax: $\Pom(p \parop c) = \Pom(p) \parop \Pom(c)$.
The denotation of a command $c$ is a subset of $\Poms(\Acts)$, also inductively defined on its syntax.
The basic commands are given by:
\begin{align*}
  \Pom(\cskip) &= \{ \{\delta\}\}\\
  \Pom(\cawrite{\mu}{x}{e}) &= \{ P \seqop \{ \awa{\mu}{x}{v} \} \mid (P, v) \in \Pom(e) \}\\
  \Pom(\cfence{\alpha}) &= \{ \{\afence\alpha\}\}\\
  \Pom(\crmw{\alpha}{a}{f}) &= \{ \{ \armw{\alpha}{a}{v}{v'}\} \mid (v, v') \in \fgraph(f) \}
\end{align*}
The only interesting clause here is for $\cawrite{\mu}{x}{e}$, where data dependency requires that the corresponding write be sequenced \textit{after} all actions performed in computing $e$.

Before we can give semantic clauses for compound commands, we must introduce the \textit{relaxed} sequential composition.
The relaxed composition of two pomsets orders actions from the first before those of the second only when required by the memory model.
To make this precise, we introduce the following predicates.
$\IsAcq(\lambda)$ holds if and only if $\lambda \in \Ar\cup\Af$ and $\Mo(\lambda) \geq \acquire$.
$\IsRel(\lambda)$ holds if and only if $\lambda \in \Aw\cup\Af$ and $\Mo(\lambda) \geq \release$.
Actions $\lambda$ and $\lambda'$ are \defin{memory-ordered}, $\Ord(\lambda, \lambda')$, if and only if $\Ide(\lambda) = \Ide(\lambda')$, $\IsAcq(\lambda)$, or $\IsRel(\lambda')$.
The \defin{relaxed sequential composition} $(P_1, {<_1}, \Phi_1) \relop (P_2, {<_2}, \Phi_2)$ of pomsets is $(P_1 \uplus P_2, {<}^+, \Phi_1 \uplus \Phi_2)$ when $P_1$ is finite, where $(i, p) < (j, q)$ if and only if $i = j$ and $p <_i q$, or $i = 1$, $j = 2$, and $\Ord(\Phi_1(p), \Phi_2(q))$, and $<^+$ is the transitive closure of $<$.
When $P_1$ is infinite, $P_1 \relop P_2 = P_1$.
Relaxed sequential composition is also associative with $\emptypset$ as its unit.

The sequencing, looping, and conditional clauses are given by:
\begin{align*}
  \Pom(c_1 \seqop c_2) &= \Pom(c_1) \relop \Pom(c_2)\\
  \Pom(\cif{b}{c_1}{c_2}) &= \left(\Pomt{b} \seqop \Pom(c_1)\right) \cup \left(\Pomf{b} \seqop \Pom(c_2)\right)\\
  \Pom(\cwhile{b}{c}) &= \left(\,\bigcup_{n=0}^\infty I^n(b, c)\right) \cup I^\omega(b,c)\\
  I^0(b,c) &= \Pomf{b}\\
  I^{n+1}(b,c) &= \Pomt{b} \seqop \left(\Pom(c) \relop I^{n}(b,c)\right)
\end{align*}
where $I^\omega(b,c)$ is the taken to be the evident infinite unfolding.
There are a few subtleties in these clauses.
In the clause for $\cif{b}{c_1}{c_2}$, we use sequential compositions because there is a control-flow dependency between the memory actions for $b$ and those for the $c_i$.
Respecting this dependency is important to eliminating ``thin-air'' behaviours.
Indeed, suppose we had used the relaxed composition instead, and consider the program $\cifthen{\earead{\relaxed}{y}=1}{\cawrite{\relaxed}{x}{1}} \parop \cifthen{\earead{\relaxed}{x}=1}{\cawrite{\relaxed}{y}{1}}$.
Then it would have a pomset of the form $\{\xymatrix@1{\ara{\relaxed}{y}{1} & \awa{\relaxed}{x}{1} & \ara{\relaxed}{x}{1} & \awa{\relaxed}{y}{1}}\}$ in its denotation, and none of the memory actions would be ordered because the locations in the boolean expressions and the commands in the conditional branches involve different locations.
One could then perform the write actions before the read actions and execute the whole program, even from a state where $y$ and $x$ are initialized to $0$.
By instead using sequential composition, the reads are sequenced before the branch's writes, and the program is not executable from this state.
In contrast, in the clause for $c_1 \seqop c_2$, we should be permitted to reorder memory accesses if the memory model allows it, and so we use the relaxed sequential composition.

The last clause is for local assignables, which can be thought of as registers.
Given a command $\clocal{n}{v}{c}$, the intention is that the assignable $n$ should be initialized to $v$ and be visible only to $c$.
Consequently, any other commands $c'$ should not be able to observe $c$'s effects on $n$, even if $n$ appears free in $c'$.
We must, however, be able to observe that $c$ did an action whenever it does an $n$-action: the program $\clocal{n}{0}{(\cwhile{\earead{\nonat}{n} = 0}{\cawrite{\nonat}{n}{0}})}$ should be non-terminating.
To satisfy these desiderata we take all of the pomsets of $c$ whose uses of the location $n$ are internally consistent and then replace all $n$-actions with no-op $\delta$ actions.
We formally accomplish this by introducing additional operations on pomsets.
To ensure internal consistency on $n$, we need to restrict our attention to $n$-actions.
The \defin{restriction} of a pomset $(P, <, \Phi)$ to a subset $L' \subseteq L$ is the pomset $\PRestr{P}{L'} = (\Phi^{-1}(L'), {<} \cap L' \times L', \Phi \restriction \Phi^{-1}(L'))$ obtained by discarding all elements whose label is not in $L'$.
To make sure they are internally consistent, we check that they are sequentially executable.
This is accomplished with a predicate $\SExec{n}{P}$ that holds if and only if $P$ is $\awa{\nonat}{n}{v}$ followed by zero or more occurrences of $\ara{\nonat}{n}{v}$, or if $P = P_1 \seqop P_2$ with $\SExec{n}{P_1}$ and $\SExec{n}{P_2}$.
This syntactic check is equivalent to the sequential executions of Section~\ref{sec:exec-interpr}.
Finally, to replace all $n$-actions by $\delta$ actions, we need a substitution operation.
The \defin{substitution} of $l$ for $L' \subseteq L$ in $P$, $[l / L']P$, is given by $(P, <_P, \Phi)$ where $\Phi(p) = l$ if $\Phi_P(p) \in L'$, and $\Phi(p) = \Phi_P(p)$ otherwise.
Combining these ingredients, we get the clause
\[
  \Pom(\clocal{n}{v}{c}) = \{ [\delta / \Acts_n]P \mid P \in \Pom(c), \SExec{n}{\{\awa{\nonat}{i}{v}\} \seqop \PRestr{P}{\Acts_n}} \}.
\]
This definition satisfies various desirable equivalences, such as
\begin{align*}
  \clocal{n}{0}{\cawrite{\nonat}{n}{42}} &\Pequiv \cskip\\
  \clocal{n}{0}{(\cwhile{\earead{\nonat}{n} = 0}{\cawrite{\nonat}{n}{0}})} &\Pequiv \cwhile{\etrue}{\cskip},
\end{align*}
where $p \Pequiv p'$ is \defin{program equivalence}, defined to hold if and only if $\Pom(p) = \Pom(p')$.

To illustrate our semantic clauses, we observe that the command
\[ \cwhile{i < 2}{(\cawrite{\nonat}{x}{\earead{\nonat}{x} + 1}\seqop \cawrite{\nonat}{i}{\earead{\nonat}{i} + 1})} \]
includes pomsets of the following form, for each $v \in V$, in its denotation:
\[{
    \xymatrix{
      & \ara{\nonat}{i}{0} \ar[r] & \awa{\nonat}{i}{1} \ar[rr] \ar[]+DR;[dr]+UL & & \ara{\nonat}{i}{1}\ar[r] & \awa{\nonat}{i}{2}\\
      \ara{\nonat}{i}{0} \ar[]+UR;[ur]+DL \ar[]+DR;[dr]+UL & & & \ara{\nonat}{i}{1} \ar[]+UR;[ur]+DL \ar[]+DR;[dr]+UL & & & \\
      & \ara{\nonat}{x}{v} \ar[r] & \awa{\nonat}{x}{v + 1} \ar[rr] & & \ara{\nonat}{x}{v + 1}\ar[r] & \awa{\nonat}{x}{v + 2}.
    }
  }
\]
In contrast, the command
\[\clocal{i}{0}{\cwhile{i < 2}{(\cawrite{\nonat}{x}{\earead{\nonat}{x} + 1}\seqop\cawrite{\nonat}{i}{\earead{\nonat}{i} + 1})}} \]
has executable pomsets of the form
\[{\xymatrix{\ara{\nonat}{x}{v} \ar[r] & \awa{\nonat}{x}{v+1}\ar[r] & \ara{\nonat}{x}{v+1} \ar[r] & \awa{\nonat}{x}{v+2}}.}
\]

\section{Executional Interpretation}
\label{sec:exec-interpr}

We give a race-detecting input-output interpretation to the abstract denotations of Section~\ref{sec:pomset-denotations}.
This interpretation serves three main purposes.
First, it gives us a notion of ``running'' the executions a pomset describes, and it tells us the initial states from which we can do so, along with the corresponding effects on state.
Second, it gives us a means of detecting which syntactic races are meaningful, and which can not occur.
For example, the program \eqref{eq:1} (page \pageref{eq:1}) has a syntactic race on the non-atomic location $x$, but this race can never occur during an execution starting from a zero-initialized state because of the synchronization via the atomic location $y$.
Finally, it allows us to rule out various pomsets assigned to commands that are not executable alone, but that are included for the sake of compositionality and that are executable in a larger environment.
Consider, for example, the pomset ${\xymatrixcolsep={1em}\xymatrix@1{\awa{\seqcons}{x}{2} \ar[r] & \ara{\seqcons}{x}{1} \ar[r] & \awa{\seqcons}{y}{1}}}$ belonging to the program $\cawrite{\seqcons}{x}{2} \seqop \cifthen{\earead{\seqcons}{x} = 1}{\cawrite{\seqcons}{y}{1}}$.
It is not executable, but it would be if we were to compose it with $\awa{\seqcons}{x}{1}$.

We use two kinds of state: proper and overdefined.
\defin{Proper states} are finite partial function from identifiers to values, in particular, elements of $\Ides \pfin V_\bot$.
We include a least element $\bot$ in the codomain to denote an unconstrained value.
Its purpose will be made clear when we define footprints of actions below.
We use the notation $\mstate{x_1 : u_1, \dotsc, x_n : u_n}$ to mean the proper state whose graph is $\{(x_1, u_1), \dotsc, (x_n, u_n)\}$.
Given proper states $\sigma$ and $\sigma'$, let $\sigma \sqsubseteq \sigma'$ if and only if for all $x \in \dom(\sigma)$, $\sigma(x) \sqsubseteq \sigma'(x)$.
The symbol $\top$ is the \defin{overdefined state}, which is the result of a race.
Let $\Sigma$ be the set of all states, ranged over by $\sigma$.

We proceed in two stages.
We first assign an executional meaning to individual memory actions.
Then, we assign an executional meaning to action pomsets.
Because we are in a weak memory setting in which a single location can be acted on concurrently, the concept of a ``global state'' is not well-defined.
Indeed, hardware features such as write buffers could cause different threads to read different values from the same location at the same time.
Instead, we use a local notion of state called a footstep.
A \defin{footstep} of an action $\lambda$ is a pair $(\sigma, \tau)$ of states, where $\sigma$ is a minimal piece of state enabling $\lambda$ to be performed, and $\tau$ describes the effect of performing $\lambda$ from $\sigma$.
The \defin{footprint} $\footp{\lambda}$ of $\lambda$ is the set of all of its footprints.
We define the footprints of memory actions as follows:
\begin{align*}
  \footp{\ara{\mu}{x}{v}} &= \{(\mstate{x:v}, \emptystate) \} & \footp{\delta} &= \{(\emptystate, \emptystate)\}\\
  \footp{\awa{\mu}{x}{v}} &= \{(\mstate{x:\bot}, \mstate{x:v}) \} & \footp{\afence\alpha} &= \{(\emptystate, \emptystate)\}\\
  \footp{\armw{\alpha}{a}{v}{v'}} &= \{(\mstate{a:v}, \mstate{a:v'})\} & &
\end{align*}
Informally, it should be clear that none of the above actions cause any allocation: whenever $(\sigma, \tau) \in \footp{\lambda}$ for some action $\lambda$, $\dom(\tau) \subseteq \dom(\sigma)$.
We use the $\bot$ value in the codomain of proper states to indicate that, though a write action $\awa{\mu}{x}{v}$ requires that the location $x$ appear in the initial state, it is ambivalent to its value.
The footprint of an action is also agnostic of the action's memory ordering tag.

We can give pomsets an analogous notion of footprint.
We will do so by recursing on the structure of the pomset, considering three principle cases: when the pomset is a single action, when the pomset can be decomposed into a pair of parallel pomsets, and when the pomset has an executable prefix.

We first specify structural conditions for when two pomsets can be run concurrently and whether doing so constitutes a race.
Because we want a total order on all $\seqcons$ actions, we cannot run two pomsets containing $\seqcons$ actions concurrently.
So we say concurrently executing $P_1$ and $P_2$ respect $\seqcons$ actions, $P_1 \rsc P_2$, if and only if only one of them performs $\seqcons$ actions, \ie, if and only if $\PRestr{P_1}{\Acts_\seqcons} = \emptyset$ or $\PRestr{P_2}{\Acts_\seqcons} = \emptyset$.
We say that pomsets $P_1$ and $P_2$ have a \defin{data race}\footnote{One could replace $\Acts_\nonat$ with $\Acts_\pi$ throughout to instead consider races between actions satisfying $\pi$.} on $n$ if $n \in \RLoc(P_1, P_2) = \bigcup_{1 \leq i \neq j  \leq 2} \Ide(P_i \restriction_{\Acts_{\nonat} \cap \Aw}) \cap \Ide(P_j \restriction_{\Acts_{\nonat}})$ and that they have a data race, $P_1 \dr P_2$, if they have one on some $n$.
Intuitively, $n \in \RLoc(P_1, P_2)$ means that $P_1$ and $P_2$ both act on $n$ with at least one of them writing to $n$.

Pomsets $P_1$ and $P_2$ are \defin{consistent}, $P_1 \co P_2$, if
\begin{inparaenum}
\item $\neg(P_1 \dr P_2)$,
\item $P_1 \penalty 100000\rsc\penalty 100000 P_2$, and
\item $\Ide(\PRestr{P_1}{\Aw}) \cap \Ide(\PRestr{P_2}{\Aw}) = \emptyset$.
\end{inparaenum}
Consistency means that there is no syntactic constraint preventing us from considering concurrent execution of $P_1$ and $P_2$.
The third condition means we do not have any write-write races between $P_1$ and $P_2$, and is required to totally order writes on a per-location basis.
In contrast, pomsets $P_1$ and $P_2$ \defin{could race}, $P_1 \rc P_2$, if
\begin{inparaenum}
\item $P_1 \dr P_2$,
\item $P_1 \rsc P_2$, and
\item $\Ide(\PRestr{P_1}{(\Aw \setminus \Acts_{\nonat})}) \cap \Ide(\PRestr{P_2}{(\Aw \setminus \Acts_{\nonat})}) = \emptyset$.
\end{inparaenum}
The third condition means that we do not have any atomic write-write races.
The intention is whenever $P_1 \rc P_2$, we should be able to regain consistency by deleting all of the data races.

Next, we need a notion of splitting a pomset into a prefix and a suffix that can sequentially be executed.
We say that a subset $Q$ of a pomset $P$ is \defin{downward-closed} if whenever $p <_P q$ and $q \in Q$, then $p \in Q$.
We write $\prefix{P_1}{P}{P_2}$ to mean that $P_1$ is a finite downward-closed subset of $P$ and that $P_2$ is the remainder of $P$.
In this case, we call $P_1$ a \defin{prefix} of $P$ and $P_2$ a \defin{suffix} of $P$.
Observe that if $P = P_1 \parop P_2$ and $P_1$ is finite, then $\prefix{P_1}{P}{P_2}$; finiteness is needed to guarantee fairness.

When executing two threads in parallel, we need only consider footsteps starting from consistent states.
We say two proper states $\sigma, \sigma' \in \Sigma$ are \defin{consistent}, $\scons{\sigma}{\sigma'}$, if $\sigma \sqcup \sigma'$ exists.
This means that for all $x \in \dom(\sigma) \cap \dom(\sigma')$, if $\sigma(x) \neq \bot$ and $\sigma'(x) \neq \bot$, then $\sigma(x) = \sigma'(x)$.
The overdefined state $\top$ is consistent with no state.
Given a set $S$ and a state $\sigma$, we let $\sigma \setminus S$ be $\top$ when $\sigma = \top$, and $\{ (x,v) \in \sigma \mid x \notin S\}$ otherwise.
Given proper states $\sigma$ and $\sigma'$, \defin{updating} $\sigma$ by $\sigma'$ gives us a new state $\supd{\sigma}{\sigma'} = \sigma' \sqcup (\sigma\setminus\dom\sigma')$.
Explicitly, $\supd{\sigma}{\sigma'}(x)$ is $\sigma'(x)$ whenever $x \in \dom(\sigma')$, and $\sigma(x)$ whenever $x \notin \dom(\sigma')$.
If $\sigma$ or $\sigma'$ is $\top$, then $\supd{\sigma}{\sigma'}$ is defined to be $\top$.
To combine the initial states of two pomsets data racing on $R \subseteq \Ides_\nonat$, we define the \defin{racy product} $\sigma_1 \otimes_R \sigma_2 = \supd{\sigma_1 \sqcup \sigma_2}{(\sigma_1 \sqcap \sigma_2) \restriction R}$, explained below.

We let the \defin{footprint} $\footp{P}$ of a pomset $P$ be inductively defined as the least set given by the following rules, which are explained below:
\begin{enumerate}
\item[\rn{Act}] If $P = \{\lambda\}$, then $(\sigma, \tau) \in \footp{P}$ for all $(\sigma, \tau) \in \footp \lambda$.
\item[\rn{Seq}] If $\prefix{P_1}{P}{P_2}$, $(\sigma_i, \tau_i) \in \footp{P_i}$, $\scons{\supd{\sigma_1}{\tau_1}}{\sigma_2}$, and $\top \notin \{ \tau_1, \tau_2 \}$,\\
  then $\scomb{\sigma_1}{\tau_1}{\sigma_2}{\tau_2} \in \footp{P}$.
\item[\rn{Par}] If $P = P_1 \parop P_2$, $P_1 \co P_2$, $(\sigma_i, \tau_i) \in \footp{P_i}$, $\scons{\sigma_1}{\sigma_2}$, and $\top \notin \{\tau_1, \tau_2\}$,\\
  then $(\sigma_1 \sqcup \sigma_2, \tau_1 \sqcup \tau_2) \in \footp{P}$.
\item[\rn{Race}] If $P = P_1 \parop P_2$, $P_1 \rc P_2$, $(\sigma_i, \tau_i) \in \footp{P_i}$, $\scons{\sigma_1}{\sigma_2}$,  and $\top \notin \{\tau_1, \tau_2\}$,\\
  then $(\sigma_1 \otimes_{\RLoc(P_1, P_2)} \sigma_2, \top) \in \footp{P}$.
\item[\rn{RaceP}] If $\prefix{P_1}{P}{P_2}$ and $(\sigma_1, \top) \in \footp{P_1}$, then $(\sigma_1, \top) \in \footp{P}$.
\item[\rn{RaceS}] If $\prefix{P_1}{P}{P_2}$, $(\sigma_i, \tau_i) \in \footp{P_i}$, $\scons{\supd{\sigma_1}{\tau_1}}{\sigma_2}$, and $\tau_2 = \top$,\\
  then $\rscomb{\sigma_1}{\tau_1}{\sigma_2} \in \footp{P}$.
\end{enumerate}
The set of \defin{executions} of a pomset $P$ is $\Exec(P) = \{ (\sigma, \supd{\sigma}{\tau}) \mid \sigma \in \Sigma, (\sigma', \tau) \in \footp{P}, \sigma' \sqsubseteq \sigma, \Ima(\sigma) \subseteq \V \}$, and the set of executions of a program $p$ is $\Exec(p) = \bigcup_{P \in \Pom(p)} \Exec(P)$.
Executions capture running programs on ``real'' states, so we require initial states to have a specific value for each location.
We say $P$ is executable from $\sigma$ if $(\sigma, \tau) \in \Exec(P)$ for some $\tau$, and $P$ is \defin{racy} if $(\sigma, \top) \in \footp{P}$ for some $\sigma$.

We explain each of the rules in turn.
The \rn{Seq} rule captures sequential execution of a prefix $P_1$ before a suffix $P_2$.
The consistency condition $\scons{\supd{\sigma_1}{\tau_1}}{\sigma_2}$ tells us that, if we start from a state satisfying $\sigma_1$ and update it with the effects $\tau_1$ of performing $P_1$, the resulting state doesn't disable the execution of $P_2$.
The state $\sigma_2 \setminus \dom\tau_1$ contributes the initial state required by $P_2$ that isn't provided by $P_1$.

The \rn{Par} rule tells us that whenever $P_1$ and $P_2$ cannot race and agree on their initial states, then we can run them in parallel with resulting effect $\tau_1 \sqcup \tau_2$.
The effect $\tau_1 \sqcup \tau_2$ is a well-defined proper state because the assumption $P_1 \co P_2$ guarantees that $P_1$ and $P_2$ do not write to the same location, \ie, that $\dom(\tau_1) \cap \dom(\tau_2) = \emptyset$.

The \rn{RaceP} rule handles races a in pomset's prefix.
The rule tells us that if $\sigma_1$ is sufficient to reach a race in $P_1$ and $P_1$ is a prefix of $P$, then $\sigma_1$ is sufficient to reach a race in $P$.
This captures the viewpoint that if we ever encounter a race, we do not need to execute the rest of the program.
The \rn{RaceS} rule deals with a race in a suffix of a pomset, and is analogous to the \rn{Seq} rule.

The \rn{Race} rule resembles the \rn{Par} rule.
The key difference is how we form the initial state.
Before considering motivating examples, we first unpack the definition of the racy product $\sigma = \sigma_1 \otimes_{\RLoc(P_1, P_2)} \sigma_2$, assuming $\scons{\sigma_1}{\sigma_2}$.
If $x \notin \dom(\sigma_2)$, then $\sigma(x) = \sigma_1(x)$, and vice-versa.
Now consider $x \in \dom(\sigma_1) \cap \dom(\sigma_2)$.
If $x \in \RLoc(P_1, P_2)$, \ie, if $P$ has a race on the non-atomic location $x$, then $\sigma(x) = \sigma_1(x) \sqcap \sigma_2(x)$, i.e., $\sigma(x) = \bot$ if any of the $\sigma_i(x)$ is $\bot$.
As we will see in the example~$Q_\mu$ below, this captures a race where the action writing to $x$ does not depend on a prior read from $x$ in order to be executable.
If $x \notin \RLoc(P_1, P_2)$, then $\sigma(x) = \sigma_1(x) \sqcup \sigma_2(x)$.
For example, $\mstate{x:0, a:1, n:2, n':3, n'':4} \otimes_{\{a,n,n'\}} \mstate{a:\penalty 1000 1, n:\penalty 1000 \bot, n' :3, n'':\bot}$ is $\mstate{x:0, a:1, n:\bot, n':3, n'':4}$

To begin with, consider the pomset $Q_\mu = \{ \xymatrix@1{ \awa{\mu}{x}{v} & \ar[l] \awa{\mu}{x}{0} \ar[r] & \ara{\mu}{x}{u} } \}$, and assume first that $\mu = \nonat$.
We argue that this pomset should be executable and racy for all values of $u$ and $v$: it has a non-atomic write of $v$ to $x$ that is not sequenced with the non-atomic read of $u$ from $x$, and so the values should not matter.
Even when $u \notin \{0, v\}$, it could be that $u$ is the value read from an intermediate hardware state caused by the writing of $x$.
Our semantics validates this desideratum: we can apply \rn{Race} to the unsequenced actions to get the footstep $(\mstate{x:\bot}, \top)$, and then using \rn{RaceS} we get that $(\mstate{x:\bot}, \top) \in \footp{Q_\nonat}$.
Now assume that $\mu \neq \nonat$, then $Q_\mu$ is not $\nonat$-racy: it has no $\nonat$ actions on which it can race.
Moreover, $Q_\mu$ has a non-empty footprint only if $u \in \{ 0, v\}$, and $\footp{Q_\mu} = \{(\mstate{x:\bot}, \mstate{x:v})\}$.

Next consider the pomset $R_\mu = \{ \xymatrix@1{ \awa{\mu}{x}{v} & \ar[l] \ara{\mu}{x}{w} & \ar[l] \awa{\mu}{x}{0} \ar[r] & \ara{\mu}{x}{u} } \}$, and assume first that $\mu = \nonat$.
We argue that this pomset should be executable only when $w = 0$: when $w \neq 0$, there is no write supplying the value $w$ required by the read, and so the data-race on $x$ should not be able to manifest itself.
However, when $w = 0$, this pomset should be racy, because we have an unsequenced non-atomic write and read from $x$.
Our semantics captures these behaviours.
For example, $\footp{\xymatrix@1{ \awa{\mu}{x}{v} & \ar[l] \ara{\mu}{x}{w}}} = \{(\mstate{x:w}, \mstate{x:v})\}$ and $\footp{\awa{\mu}{x}{u}} = \{(\mstate{x:u},\emptystate)\}$, and the states $\mstate{x:w}$ and $\mstate{x:u}$ are consistent only if $w = u$.
When this is the case, we get $(\mstate{x:w}, \top)$ as the sole footprint for the parallel composition.
But we can apply the \rn{RaceS} rule only when $\scons{\supd{x:\bot}{x:0}}{\mstate{x:w}}$, which holds only when $w = 0$.
Decomposing the pomset differently provides the same constraint.
In contrast, when $\mu \neq \nonat$, $R_\mu$ is not racy and it is executable only when $w = 0$ and $u \in \{0,v\}$.

Finally, consider the pomset $S = \{ \xymatrix@1{ \awa{\nonat}{x}{v} \ar[r] & \awa{\release}{y}{1} & \ara{\acquire}{y}{1} \ar[r] & \ara{\nonat}{x}{u} } \}$.
We get the footsteps $(\mstate{x:u, y:1}, \mstate{x:v, y:1})$ and $(\mstate{x:\bot, y:1}, \top)$ for all $u$ and $v$.
The first footstep corresponds to a sequential execution with all the reads before the writes.
The second footstep corresponds to $y$ reading $1$ from the initial state, and then the program racing on $x$.
When $u = v$, we also get the footstep $(\mstate{x:\bot, y:\bot}, \mstate{x:v, y:1})$, which describes an execution where all reads are performed after all writes.

We show that our definition of execution validates various other litmus tests in the way we expect.
The \textsc{sb} (store buffering) test is the simplest example of behaviour that is not sequentially consistent.
When $\alpha \neq \seqcons$, we can execute the pomset $\{ \xymatrix@1{ \awa{\alpha}{x}{1} \ar[r] & \ara{\alpha}{y}{0} & \awa{\alpha}{y}{1} \ar[r] & \ara{\alpha}{x}{0} } \}$ starting from the state $\mstate{x:0, y:0}$ by using the \rn{Par} rule.

We can also validate the \textsc{iriw} (independent reads of independent writes) test.
Indeed, we can execute the pomset
\[
  \xymatrix{ \awa{\release}{x}{1} & \ara{\acquire}{y}{1} \ar[r] & \ara{\acquire}{x}{0} & \awa{\release}{y}{1} & \ara{\acquire}{x}{1} \ar[r] & \ara{\acquire}{y}{0}
  }
\]
starting from the initial state $\mstate{x:0, y:0}$ by splitting the pomset down the middle and applying the \rn{Par} rule.
This shows that we are weaker than the TSO memory model, because we do not impose a total order on all writes.
Though we can read writes to different locations in different orders, the $\co$ restriction on the \rn{Par} rule ensures that we have a per-location total order on writes.
Then all threads see the writes to the same location in the same order, \ie, we guarantee coherence.

\section{Related Work}
\label{sec:related-works}

Most work on formalizing weak memory models so far uses ``execution graphs'' or ``candidate executions'', in which nodes are labeled with actions and there are multiple kinds of edge, usually characterized as \textit{po} (program order), \textit{rf} (reads-from), \textit{mo} (modification order), and so on.
There is a substantial body of well-established research in this vein, including \cite{batty13,batty12,boehm08,sarkar09,sewell10}.
In work aiming to formalize C/C++11 such as \cite{batty11}, axioms are imposed to rule out candidate executions involving undesirable cycles of composite edges, primarily to avoid issues with so-called thin-air reads, and to ensure that each read action is justified by a suitable write.
It has proven to be difficult to strike the right balance between ruling out bad cases while still allowing intended behaviours.

Our memory model is heavily based on the RC11 (``Repaired C11'') model presented by Lahav~\ea~\cite{Lahav:2017:RSC:3062341.3062352}.
The RC11 model repairs compilation to Power by providing a better semantics for SC accesses.
Like RC11, our model differs from C11 by not treating races between atomic accesses as undefined behaviour.
Our approach to eliminating ``thin-air'' behaviour is similar to theirs: we prohibit violations of data and control-flow dependency, which amounts to prohibiting cycles in the execution (``\texttt{hb}'') order between reads and their corresponding writes.

In prior work~\cite{Kavanagh17}, we give a semantics to the SPARC TSO weak memory model using pomset denotations and executions, and buffered state.
The semantics developed in this paper is a significant step forward toward greater generality and wider applicability.
Rather than attempt to model out-of-order execution by explicitly modelling buffers, both in pomset generation and in execution, we use the relaxed composition operator and leverage pomset structure.
To enforce a total order on writes during execution, TSO executions were parametrized with linearizations of writes.
In contrast, C11 executions totally order $\seqcons$ actions via the $\co$ relation, thereby reducing technical overhead.

Jeffrey and Riely~\cite{Jeffrey:2016:TAR:2933575.2934536} and Castellan~\cite{Castellan16} give a denotational semantics using event structures and exploit game-theoretic ideas in formulating a notion of execution.
We prefer to work with a set of pomsets, obtained by taking account of possible relaxations and executions, rather than using a single structure combining these possibilities into one object; it seems less cumbersome to work with sets-of-pomsets.

\section{Conclusion}
\label{sec:conclusion}

Our extension to incorporate the wider range of C11-style memory orderings required significant technical development in order to produce a denotational account that is faithful to operational intuitions, and improves on the foundations laid by our prior denotational account of TSO.
A key new idea presented here is the relaxed sequential composition for pomsets, a form of sequential composition that takes account of the memory model's support for reordering of memory actions.
We are careful to avoid reordering of actions for which there is a control-flow dependency, arguing that this eliminates a major class of thin-air behaviours.
We discussed a selection of ``litmus test'' examples to show that our semantics and our notion of pomset execution yield results consistent with the literature.
We plan a more comprehensive cataloguing of litmus tests to solidify this claim.
Indeed, we expect to prove a pomset analogue of the DRF-SC property, which provides programmers a sufficient condition for ensuring their programs are not executionally racy.

\bibliographystyle{entcs}
\bibliography{mfps34}

\end{document}